\documentstyle[12pt]{article}
\font\srm=cmr9

\def\beq{\begin{equation}} \def\eeq{\end{equation}}

  \def\fe{f^{\rm e}}

\def\be{\begin{equation}}
\def\fe{\end{equation}}

\begin{document}

\title{\bf Relativistic solution of Iordanskii problem\\
in multi-constituent superfluid mechanics}

\author { {\bf B. Carter, D. Langlois, R. Prix} \\ \hskip 1 cm\\
D\'epartement d'Astrophysique Relativiste et de Cosmologie,\\
Centre National de la Recherche Scientifique, \\Observatoire de
Paris, 92195 Meudon, France.}

\date{\it 25 March, 2000}

\maketitle 

\vskip 0.4  cm
\noindent 
{\bf Abstract.\ } Flow past a line vortex in a simple perfect fluid or
superfluid gives rise to a transverse Magnus force that is given by
the well known Joukowski lift formula. The problem of generalising
this to multiconstituent superfluid models has been controversial since 
it was originally posed by the work of Iordanski in the  
context of the Landau 2-constituent model for $^4\!$He at finite 
temperature. The present work deals not just with this particular 
case but with the generic category of perfect multiconstituent models 
including the kind proposed for a mixture of $^4\!$He and $^3\!$He by 
Andreev and Bashkin.  It is shown here (using a relativistic approach) 
that each constituent will provide a contribution proportional to the 
product of the corresponding momentum circulation integral with the 
associated asymptotic current density.

\section{Introduction}

For a simple perfect fluid with asympotically uniform density
$\overline\rho$ say, the Magnus effect of a uniform background 
flow with relative velocity $v^i$ say in the rest frame of a vortex in 
the direction of a 3 dimensional unit vector $\ell^i$ results in
a force per unit length given by the well known (non-relativistic) 
Joukowsky formula as
\be {\cal F}_i=\kappa\bar\rho\,\varepsilon_{ijk}\ell^j\overline v^k\, ,
\label{0}\fe
where $\kappa$ is the relevant velocity circulation integral.

The question raised by Iordanskii~\cite{Iord66} of how this formula 
should be generalisd to the case of Landau's 2 constituent model for 
superfluid $^4\!$He at finite temperature has been a subject of 
controversy: the most widely accepted~\cite{KopninVP95,Stone99} 
prescription is that of Sonin~\cite{Sonin76,Sonin97}, but various 
alternatives have been proposed by other 
authors~\cite{BDVinen83,Thoulessetal96,WThouless98}. 
The present work clarifies the issue by demonstrating the existence of 
an elegant generalisation of the Joukowsky formula (\ref{0}) for an 
extensive class of perfect multiconstituent fluid models, including, as 
well as the Landau model, the Andreev Bashkin model~\cite{AndreevB76} 
for a mixture of superfluid $^4\!$He with (``normal'') $^3\!$He. 
(However our analysis does not cover the more complicated 
subject~\cite{KopninVP95} of superfluid $^3\!$He). 

For technical convenience (not to mention the consideration that
it is more accurate in contexts such as that of neutron star matter) 
the work is carried out using a (special) relativistic formulation . 

\section{Perfect multiconstituent fluid dynamics}

As well as including the original Landau model in the (not so simple) 
Galilean limit~\cite{CK94}, two independent lines of early development 
of relativistic (and thus technically simpler) perfect multiconstituent 
fluid theory -- using currents~\cite{C79,C85} and momenta~\cite{LK82} 
respectively as independent variables in (suitably constrained) 
variational formulations -- were subsequently shown to be entirely 
equivalent\cite{CK92}: the independent momentum covectors 
$\mu^{_{\rm X}}_{\, \nu}$ of the latter aproach are identifiable 
just as the dynamical conjugates of the independent conserved 
current vectors $n_{_{\rm X}}^{\,\nu}$ on which the former approach 
was based. In the latter approach, which is the most convenient for 
our present purpose, the fundamental equation of state characterising 
a particular perfect multiconstituent fluid model is given by 
specifying the dependence of the relevant generalised pressure 
function $\Psi$ on the independent momentum covectors  
$\mu^{_{\rm X}}_{\, \nu}$, and the associated current vectors 
$n_{_{\rm X}}^{\,\nu}$ are then obtained as the corresponding
partial derivatives in the infinitesimal variation formula
\be \delta\Psi=-\sum_{^{\rm X}} n_{_{\rm X}}^\nu \delta
\mu^{_{\rm X}}_{\,\nu}\, .\label{14}\fe
For such a model, the complete set of equations of motion consists
just of a set of particle and vorticity conservation laws of the form
\be \nabla_{\!\nu}n_{_{\rm X}}^{\,\nu}=0\, ,\hskip 1 cm
n_{_{\rm X}}^\nu w_{\nu\sigma}=0\, .\label{15}\fe
where, for any particular constituent with label {\srm X} the
corresponding generalised vorticity 2-form is defined as the exterior
derivative
\be w^{_{\rm X}}_{\,\nu\sigma}=\nabla_{\!\nu}
\pi^{_{\rm X}}_{\,\sigma}-\nabla_{\!\sigma}
\pi^{_{\rm X}}_{\,\nu}\, ,\hskip 1 cm
\pi^{_{\rm X}}_{\,\nu}=\mu^{_{\rm X}}_{\,\nu}+e^{_{\rm X}}A_\nu
\label{16}\fe
of the generalised momentum covector $\pi^{_{\rm X}}_{\,\nu}$,
whose specification\cite{CL98} allows for the possibility of coupling to
an electromagnetic field
\be F_{\nu\sigma}=\nabla_{\!\nu}A_\sigma
-\nabla_{\!\sigma}A_\nu\, .\hskip 1 cm \nabla_{\!\sigma}F^{\sigma\nu}=
4\pi J^\nu\, ,\label{18}\fe
with current and electromagnetic stress tensor given by
\be J^\nu= \sum_{^{\rm X}} e^{_{\rm X}} n_{_{\rm X}}^{\,\nu}
 \, ,\hskip 1 cm
 T_{\rm F\, \nu}^{\,\sigma}={1\over 4\pi}\big( F^{\mu\nu}
F_{\mu\sigma}-{1\over 4} F^{\mu\rho}F_{\mu\rho} g^\sigma_{\, \nu}\big)
\, ,\label{19}\fe
where $e^{_{\rm X}}$ is the electric charge, if any, per particle
of the {\srm X} th species.

In such a model a (flat space) conservation law of the usual form 
\be \nabla_{\!\nu}T^\nu_{\ \sigma}=0\, ,\label{20}\fe
will be satisfied by the relevant total
stress energy tensor, which takes the form
\be T^\sigma_{\ \nu}=\sum_{^{\rm X}} n_{_{\rm X}}^\sigma 
\mu^{_{\rm X}}_{\,\nu}+\Psi g^\sigma_{\,\nu}+
T_{\rm F\, \nu}^{\,\sigma} = \sum_{^{\rm X}} n_{_{\rm X}}^\sigma
\pi^{_{\rm X}}_{\,\nu}+\Psi g^\sigma_{\,\nu}-J^\sigma A_\nu+
T_{\rm F\, \nu}^{\,\sigma}\, .
\label{21}\fe

The transport law (\ref{15}) for the vorticities is such that they will 
remain zero if they are zero initially. Thus we shall have
\be w^{_{\rm X}}_{\,\nu\sigma}=0\, ,\label{22}\fe
not just for cases of superfluidity or superconductivity (i.e. cases for 
which the momentum covector is the gradient of a condensate phase scalar)
but even for ``normal'' constituents in configurations of the
kind to be considered here, in which a perturbing vortex moves
through an {\it asymptotically uniform} medium characterised by vanishing
of the asymptotic background value (indicated here by an overhead bar)
not just of the current (as is necessary for uniformity) but also of 
the electromagnetic field, and (in an appropriate gauge) of its vector 
potential,
\be \overline J{^\nu}=0\, \hskip 1 cm \overline F_{\nu\sigma}=0\, ,
\hskip 1 cm \overline A_\sigma=0\, .\label{23}\fe
This must necessarily be the case (the Meissner effect), with the
implication that the uniform background value of the stress energy 
density tensor will be given simply by
\be \overline T{^\sigma}_{\!\nu}=\sum_{^{\rm X}} 
\overline n_{_{\rm X}}^\sigma  \overline \pi^{_{\rm X}}_{\,\nu}+
\overline \Psi g^\sigma_{\,\nu}\, ,\label{24}\fe
whenever even just a single one of the uniform background constituents is 
superconducting (since $e^{_{\rm X}} F_{\nu\sigma}=$ $\nabla_{\!\nu}
\pi^{_{\rm X}}_{\,\sigma}-\nabla_{\!\sigma}
\pi^{_{\rm X}}_{\,\nu}-w^{_{\rm X}}_{\,\nu\sigma}$).

\section{Specification of lift force on vortex}

The subject of this investigation is an asymptotically uniform
vortex configuration that is stationary with respect to a rest frame 
characterised by a uniform timelike unit symmetry generating vector 
field $k^\mu$ say, and that is aligned in the direction of a uniform 
orthogonal spacelike unit symmetry generating vector field $\ell^\mu$ in 
a flat background spacetime using Minkowski coordinates. Provided 
the corresponding conditions of stationarity and longitudinal
symmetry apply to $F_{\nu\sigma}$ not just outside but
even within the vortex core region, they will be applicable to the 
potential $A_\sigma$ in a suitable gauge, and hence also, not just to
the gauge independent covectors $\mu^{_{\rm X}}_\nu$ but to the
corresponding generalised momenta $\pi^{_{\rm X}}_\nu$ as well.
In view of the vanishing (\ref{22}) of the vorticity vectors
(\ref{16}), the condition that the momentum covectors should 
be invariant with respect to the action of the the uniform
symmetry generating vector fields $k^\nu$ and $\ell^\nu$ can
be seen to imply the uniformity of corresponding sets of
generalised Bernouilli constants,
\be k^\nu\pi^{_{\rm X}}_{\,\nu}=k^\nu\overline\pi^{_{\rm X}}_{\,\nu}
\, ,\hskip 1 cm  \ell^\nu\pi^{_{\rm X}}_{\,\nu}=
\ell^\nu\overline \pi^{_{\rm X}}_{\,\nu}\, .\label{30}\fe
 
The force per unit length, ${\cal F}_\nu$ acting on such a 
stationary longitudinally invariant vortex can be evaluated
as the integral round a circuit $s$ say surrounding the vortex
in an orthogonal 2-plane in the form
\be {\cal F}_\nu =\oint f_\nu ds \, ,\label{31}\fe
where $ds$ is the proper distance element given by $ds^2=g_{\nu\sigma}
dx^\nu dx^\sigma$ and $f_\nu$ is the local force density that is
given by
\be f_\nu = \nu_\sigma T^\sigma_{\ \nu}\, ,\label{32}\fe
in terms of the unit normal covector $\nu_\sigma$ which will be given 
in terms of the antisymmetric background measure tensor
$\varepsilon_{\lambda\mu\nu\sigma}$ by
\be  \nu_\sigma ds=\,^\star\!\varepsilon_{\sigma\nu} dx^\nu
\, ,\hskip  1 cm ^\star\!\varepsilon_{\sigma\nu}=\ell^\rho
\varepsilon_{\rho\sigma\nu}\, ,\hskip 1 cm \varepsilon_{\rho\sigma\nu}
=k^\mu\varepsilon_{\mu\rho\sigma\nu}\, .\label{33}\fe

\section{Generalised Joukowski theorem}

It is to be observed that, as a consequence of the conservation law
(\ref{20}), it makes no difference what circuit is employed for 
evaluating ${\cal F}_\nu$. We are thus allowed to choose a circuit 
sufficiently far out for reliability of our smoothed fluid 
description (whose physical validity might be questionable near the 
core) to be ensured, and also for the deviation from the uniform 
background value $\overline T{^\sigma}_{\!\nu}$ to be evaluated as 
a linear perturbation:
\be T^\sigma_{\ \nu}-\overline T{^\sigma}_{\!\nu}=\delta
 T^\sigma_{\ \nu}+O\{\delta^2\}\, .\label{34}\fe
Since the force integral for the unperturbed uniform background
must evidently vanish, $\overline {\cal F}_\nu=0$ by symmetry,
the corresponding value in the presence of the vortex will be
given by 
\be {\cal F}_\nu=\delta {\cal F}_\nu+ O\{\delta^2\}\, .\label{35}\fe

Using (\ref{14}) and (\ref{24}) it can immediately
be seen that the required first order variation will be given by
\be \delta T^\sigma_{\ \nu}=\sum_{^{\rm X}}\big(\overline 
\pi^{_{\rm X}}_{\,\nu}\delta  n_{_{\rm X}}^{\,\sigma}+
\overline n_{_{\rm X}}^{\,\sigma} 
\delta \pi^{_{\rm X}}_{\,\nu}- g^\sigma_{\,\nu}
\overline n_{_{\rm X}}^{\,\rho}\delta \pi^{_{\rm X}}_{\,\rho}\big)
\, .\label{36}\fe
Using the decomposition of the 4-dimensional spacetime metric in the 
form $ g^\nu_{\,\sigma}=\eta^\nu_{\,\sigma}+\perp^{\!\nu}_{\sigma}$
as the sum of the (rank 2) operators of projection respectively
parallel to and orthogonal to the vortex given by
$\eta^\nu_{\,\sigma}= -k^\nu k_\sigma+\ell^\nu\ell_\sigma$, and 
$\perp^{\!\nu}_{\sigma}=\,^\star\!\varepsilon^{\mu\nu}
\,^\star\!\varepsilon_{\mu\sigma}$ and using the possibility
of taking the Bernouilli constants (produced by
parallel projection) outside the integration, (\ref{36}) provides
a result expressible simply as
\be \delta {\cal F}_{ \nu}=\sum_{^{\rm X}}\big(
 \overline  n_{_{\rm X}}^{\,\sigma}\,
^\star\!\varepsilon_{\sigma\nu}\delta {\cal C}^{_{\rm X}}
+\overline \pi^{_{\rm X}}_{\,\nu}\delta D_{_{\rm X}}\big)
\, ,\label{40}\fe
where for each species {\srm X} the corresponding momentum 
circulation integral  ${\cal C}^{_{\rm X}}$ and current outflux 
integral $D_{_{\rm X}}$ are defined by
\be {\cal C}^{_{\rm X}}=\oint \pi^{_{\rm X}}_{\,\nu} dx^\nu
\, ,\hskip 1 cm D_{_{\rm X}}=\oint 
n_{_{\rm X}}^{\,\sigma}\nu_\sigma ds\, .\label{41}\fe
The irrotationality condition (\ref{22}) ensures that 
${\cal C}^{_{\rm X}}$ is independent of the choice of circuit, and
the current conservation law (\ref{15}) ensures that the same
will apply to $D_{_{\rm X}}$, which furthermore will simply
vanish,  $D_{_{\rm X}}=0$, provided there is no current creation
in the vortex core. Thus by the fact that the uniform background
value of the circulation integrals must also vanish, $\overline
{\cal C}{^{_{\rm X}}}=0$, and by taking the limit in which the
circuit is taken to a very large distance outside, one obtains
an exact net force formula of the simple form
\be {\cal F}_{ \nu}=\,^\star\!\varepsilon_{\sigma\nu}\sum_{^{\rm X}}
{\cal C}^{_{\rm X}} \overline n_{_{\rm X}}^{\,\sigma}
\, .\label{42}\fe
This result is the required (relativistic, multiconstituent)
generalisation of Joukowsky's well known formula (\ref{0}) for the 
single constituent case. What it means is that each constituent 
contributes an amount proportional to, but orthogonal to, its 
asymptotic current vector, with a coefficient  given by the 
corresponding momentum circulation integral.

\section{Application to the Landau model}

The particular example that motivated this work is that of
superfluid $^4\!$He at finite temperature, as described by
the Landau model in terms of just two constituents with 
conserved 3-dimensional current
densities $n_{_\alpha}^{\,i}=n_{_\alpha}v_{_\alpha}^{\,i}$ and 
$n_{_\beta}^{\,i}=n_{_\beta}v_{_\beta}^{\,i}$
of which the first represents Helium atoms, i.e. ``dressed'' alpha
particles, characterised by a ``rest mass'' $m_{_\alpha}$, and
the second represents units of entropy, characterised by
a vanishing rest mass $m_{_\beta}=0$. As in the less specialised
case of the 2 constituent (zero temperature) limit of
the Andreev Bashkin model for which the second constituent is
$^3\!$He  with non vanishing rest mass, $m_{_\beta}\simeq 
3 m_{_\alpha}/4$, the Newtonian limit description can be formulated
in terms of a total mass density and 3 dimensional mass current
\be \rho= \rho_{_\alpha}+\rho_{_\beta}\, ,\hskip 1 cm 
\rho^i=\rho_{_\alpha} v_{_\alpha}^{\,i}
+\rho_{_\beta}v_{_\beta}^{\,i}\, ,\label{46}\fe
where $\rho_{_\alpha}=m_{_\alpha}n_{_\alpha}$ and 
$\rho_{_\beta}=m_{_\beta}n_{_\beta}$,
so that the latter vanishes in the particular case of the Landau model. 
The total mass current is identifiable with the total momentum density
$\rho_i=n_{_\alpha}\mu^{_\alpha}_{\, i}+n_{_\beta}\mu^{_\beta}_{\, i}$,
in which, due to the effect of ``entrainment'' (which is describable 
in terms of ``effective masses'' that are different from the bare 
masses) the vanishing of the second contribution to the mass current 
does not imply absence of the second momentum contribution 
proportional to $\mu^{_\beta}_{\,i}$.

The pseudo-velocity $v_{_{\rm S}}^{\,i}$ that is commonly referred to as 
the ``superfluid velocity'' is defineable by $v_{_{\rm S}i} 
=m_{_\alpha}^{-1}\mu^{_\alpha}_{\,i}$. In the Landau case (unlike 
the generic Andreev Bashkin case) it is not possible to define an 
analogous pseudo velocity for the other consituent, because of the 
vanishing of $m_{_\beta}$, and the quantity commonly denoted 
as $v_{_{\rm N}}^{\,i}$ and known as the ``normal'' velocity is simply 
identifiable with the velocity of the entropy current, i.e. 
$v_{_{\rm N}}^{\,i}=v_{_\beta}^{\,i}$. In a mass and momentum 
decomposition of the commonly used (effectively ``mongrel'') form 
\be \rho=\rho_{_{\rm S}} +\rho_{_{\rm N}}\, ,\hskip 1 cm \rho^i=
\rho_{_{\rm S}} v_{_{\rm S}}^{\,i}+\rho_{_{\rm N}} v_{_{\rm N}}^{\,i}
\, ,\label{47}\fe
the coefficients $\rho_{_{\rm S}}$ and  $\rho_{_{\rm N}}$ must
not be confused with $\rho_{_\alpha}$ and  $\rho_{_\beta}$
(of which the latter is zero in the Landau case characterised by
by $\rho=\rho_{_\alpha}$). In the Landau case, as well as in the 
generic Andreev Bashkin case, there are two independently conserved 
momentum circulation integrals,
\be {\cal C}^{_\alpha}=\oint \mu^{_\alpha}_{\, i}\, dx^i\, ,\hskip
1 cm {\cal C}^{_\beta}=\oint \mu^{_\beta}_{\, i}\, dx^i
\, ,\label{50}\fe
of which, by the superfluidity property, the former is quantised,
${\cal C}^{_\alpha}=h$. Since $m_{_\alpha}\neq 0$ we can write
\be {\cal C}^{_\alpha}=m_{_\alpha}\kappa_{_{\rm S}} \, ,\hskip 1 cm
{\cal C}^{_\beta}=m_{_\beta}\kappa_{_{\rm S}}
+\oint{\rho_{_{\rm N}}\over n_{_\beta}}
(v_{_{\rm N}i}-v_{_{\rm S}i})dx^i \, , \hskip 1 cm \kappa_{_{\rm S}} 
= \oint v_{_{\rm S} i}\, dx^i \, ,\label{51}\fe 
in terms of the pseudo-velocity circulation integral 
$\kappa_{_{\rm S}}$, which has no ``normal'' analogue in the Landau 
case, because of the vanishing of $m^{_\beta}$.
Thus (in the rest frame of the vortex) using the notation 
$\,^\star\!\varepsilon_{ij}=\varepsilon_{ijk}\ell^k$,
one finally obtains a non-relativistic force formula of the form
\be {\cal F}_i=\,^\star\!\varepsilon_{ij}\big(\kappa_{_{\rm S}}
\overline\rho_{_\alpha}\overline v_{_\alpha}^{\,j}+{\cal C}^{_\beta}
\overline n_{_\beta}\overline v_{_\beta}^{\,j}\big)
=\kappa_{_{\rm S}}\overline\rho_{_{\rm S}}\,^\star\!\varepsilon_{ij}
\overline v_{_{\rm S}}^{\,j}+{\cal F}_{_{\rm I} i} \, ,\label{52}\fe
where in this last version the first term is what is commonly
referred to as the ``superfluid Magnus force'' contribution,
while the remaining ``Iordanskii'' correction term is found to
be given by
\be{\cal F}_{_{\rm I} i }=\big({\cal C}^{_\beta}\overline n_{_\beta}
+\kappa_{_{\rm S}}(
\overline\rho_{_{\rm N}}-\overline\rho_{_\beta})\big)
\,^\star\!\varepsilon_{ij}\overline v_{_{\rm N}}^{\,j} 
\, .\label{53}\fe
The third term in this expression is needed for the generic
Andreev Bashkin case, but drops out for the special Landau case 
characterised by $\rho_{_\beta}=0$.

If, as well as setting $\rho_{_\beta}$ to zero, one adopts the
 plausible supposition that the ``normal'' circulation will
vanish, ${\cal C}^{_\beta}=0$, then the first term also
drops out so that our formula will reduce to a form 
that is in exact agreement with the result that was derived by 
Sonin\cite{Sonin76,Sonin97} and confirmed, on the basis of a more 
rigorous microscopic analysis of phonon dynamics, 
by Stone\cite{Stone99}. 

This widely accepted conclusion has however been vigorously contested 
by Thouless and coworkers~\cite{Thoulessetal96,WThouless98} who have used 
a more sophisticated -- though not obviously more reliable -- kind of 
microscopic analysis to argue that the Iordanskii force term 
${\cal F}_{_{\rm I} i }$ vanishes, leaving just the purely 
``superfluid'' term (namely $^\star\!\varepsilon_{ij}\,\kappa_{_{\rm S}}
\overline \rho_{_{\rm S}}\, \overline v_{_{\rm S}}^{\,j}$)
in (\ref{52}). As {\it prima facie} evidence in favour of this dissident 
conclusion, it is to be observed that in the limit when there is
no relative flow at all (i.e. $\overline v_{_{\rm S}}^{\,i}
=\overline v_{_{\rm N}}^{\,i}=0$) then -- as a requirement for
compatibility with strict stationarity -- the  long term effect of the 
small ``normal'' viscosity contribution that was neglected in the
preceeding analysis will impose a rigidity condition to the effect
that $ v_{_{\rm N}}^{\,i}=0$ throughout. This imperative entails 
small deviations from strict irrotationality of the normal constituent 
except in the incompressible case for which the ratio 
$\rho_{_{\rm N}}/n_{_\beta}$ is exactly uniform, and it ensures 
in any case by (\ref{51}) that the normal momentum circulation round 
a circuit at large distance will be given by the formula 
$  {\cal C}^{_\beta}= \big(m_{_\beta}-\overline \rho_{_{\rm N}}/
\overline n_{_\beta}\big)\kappa_{_{\rm S}}$ whose substitution
in (\ref{53}) does indeed give, ${\cal F}_{_{\rm I} i }=0$. However 
this simple counter argument is inconclusive because -- as shown every 
time an ordinary light aircraft takes off -- the stationary circulation
value due to the long term effect of slight deviations from strictly
inviscid behaviour will change as a function of the relative flow 
velocity. 

To sum up, the present work shows how the Iordanskii force is given 
simply as a function of the (in the inviscid limit conserved) ``normal'' 
momentum circulation integral  ${\cal C}^{_\beta}$, but the issue of the
appropriate value for this parameter in a realistic steady flow
confguration is beyond the scope of a perfectly conducting fluid
treatment such as is provided here. Experience with the analogous
aerofoil problem in the context of aircraft engineering suggests
that the final resolution of this issue may involve subtleties
that have have eluded even the most sophisticated analysis available
so far.

We wish to thank Uwe Fischer, Michael Stone, and Grigori Volovik
for instructive conversations.

\end{document}